# Effect of CSMA/CD on Self-Similarity of Network Traffic

Altyeb Altaher Altyeb
Department of Computer Science
International University of Africa
altypaltaher@uofk.edu

**Abstract** - *It is now well known that Internet traffic exhibits self-similarity, which cannot be described by traditional Markovian models such as the Poisson process. The causes of self-similarity of network traffic must be identified because understanding the nature of network traffic is critical in order to properly design and implement computer networks and network services like the World Wide Web. While some researchers have argued self similarity is generated by the typical applications or caused by Transport layer Protocols, it is also possible that the CSMA/CD protocol may cause or at least contribute to this phenomenon. In this paper, we use NS simulator to study the effect of CSMA/CD Exponential Backoff retransmission algorithm on Traffic Self similarity.*

## I. INTRODUCTION

Recent studies have shown the presence of self-similarity in Ethernet LAN traffic [9], World Wide Web traffic [7], Wide Area Network traffic [8], etc. The issue of self-similarity has also been addressed in various studies from many different aspects including its effect on network performance [5], modeling techniques [5, 1], and causes of the appearance of self-similarity [4,2].

Since the pioneering work on self-similarity of network traffic by Leland et. al., many studies have attempted to determine the cause of this phenomenon. Initial efforts focused on application factors. For example, Crovella and Bestavros [7] investigated the cause of self-similarity by focusing on the variability in the size of the documents transferred and the inter-request time. They proposed that the heavy-tailed distribution of file size and "user thinks time" might potentially be the cause of self-similarity found in web traffic. Alternatively, few studies have considered the possibility that underlying network protocols such as TCP may cause or exacerbate the phenomenon [4,3, 6].

In this work, the network simulator (NS-2 ) used to study the effect of the CSMA/CD protocol on network traffic and its contribution to the self-similarity phenomenon . The developed model in this work assumes finite load and uses the IEEE 802.3(Media Access Protocol) MAC protocol that employs the Carrier Sense Multiple Access with Collision Detection (CSMA/CD) protocol.

This paper is organized as follows. Section II describes the network model and network configuration. Results are presented and discussed In section III.Concluding remarks are given in section IV.

## II. The Network Model

### A. Network Model

Figure 1 illustrates the network model used in this work , the model is a server-client model, in which 32 clients are connected to two servers via a single link with a bandwidth of 10_Mbps.Network simulation package NS is used here ,with some modifications in codes. In the used model, each client requests a file transfer to a randomly selected server, and the server sends a file back as a series of fixed size (1K byte) packets. The following parameters are considered for the Ethernet: Max Propagation Delay 950_ns, Jam time after Collision 3.2 ms, Slot Size 51.2 ms , and Inter-frame Delay 9.6 ms

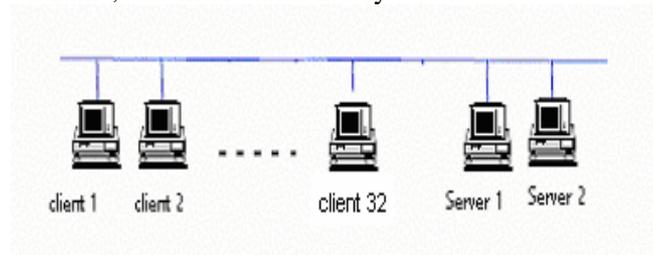

Figure 1: Network model

### B. Signs of Traffic Self-Similarity

The above network has been configured to behave as sporadically congested network. The network alternates between congested and uncontested periods in order to examine the effect of CSMA/CD Exponential Backoff retransmission algorithm on Traffic Self similarity.

Throughout this paper. Rescaled Adjusted Range Statistic (also called the R/S statistic) is used to measure the degree of Traffic self-similarity.

Hurst Parameter ($0.5 < H < 1.0$) which measure the degree of self-similarity calculated as follow, for a given set of observations $x_1, x_2, \ldots, x_n$ with sample mean $\bar{x}(n)$ and sample variance $s^2(n)$ the rescaled adjusted range or R/S statistic is given by

R(n)/S(n) = (1/s(n))(max(0, $x_1, x_2, \ldots, x_n$) − min(0, $x_1, x_2, \ldots, x_n$))

With $w_k = x_1 + x_2 + \ldots + x_k - k\bar{x}$, k = 1, 2, …, n. Hurst found that many naturally occurring time series well represented by the relation

$$E[R(n)/S(n)] \approx cn^H, n \to \infty$$

## C. CSMA/CD Exponential Backoff retransmission

Ethernet uses a Binary Exponential Backoff Algorithm to handle collisions. The algorithm is designed to dynamically adapt to the number of stations that are trying to send.

The binary exponential backoff algorithm dynamically adapts itself to the number of stations and their level of activity. It is designed to avoid excessively long average waits after a collision (e.g. if a randomization interval of 1023 was used, then the chance of two stations colliding is negligible, however, the average wait after a collision is excessive). The approach is also designed to avoid deadlock.

## III. The Results

Figure 2 shows time series plots of network traffic measured in bytes per time unit, as a function of time. The figure shows plots which span three orders of magnitude in time scale and three different ranges of maximum number of retransmission. The time units used vary from 5 second in the left most column to 20 sec in the right most column. The three rows show how traffic varies when the maximum number of retransmissions equal 3, 6, 9 respectively.

It worth to notice that smaller maximum number of retransmission generates greater traffic self similarity because After the collision, each station waits slot times in range [0, $2^i - 1$] (where i is the number of collision that have occurred in a row) before trying again. As a result, it was found that the maximum number of retransmission has great effect on the traffic self similarity.

Although Figure 2 gives some evidence of the difference in self-similarity traffic as a function of maximum number of retransmission, a quantitative measure of self-similarity is obtained by using the Hurst parameter H which expresses the speed of decay of a time series' autocorrelation function. Hurst parameter which measure the degree of self-similarity calculated by the relation

$$E[R(n)/S(n)] \approx cn^H, n \to \infty$$

| Traffic Self similarity | 0.730 | 0.720 | 0.719 |
|---|---|---|---|
| Maximum No of Retransmission | 3 | 6 | 9 |

Table 1. Hurst parameter estimates bsed on (R/S) for Maximum Number of retransmission varying from 3 to 9.

Table 1 shows H-estimates based on R/S method for the network model illustrated in figure 1. The table shows H as a function of the maximum number of retransmission equal to 3; 6 and 9. It was notice that Hurst parameter vary with the maximum number of retransmission. The higher traffic self similarity found at smallest maximum number of retransmission.

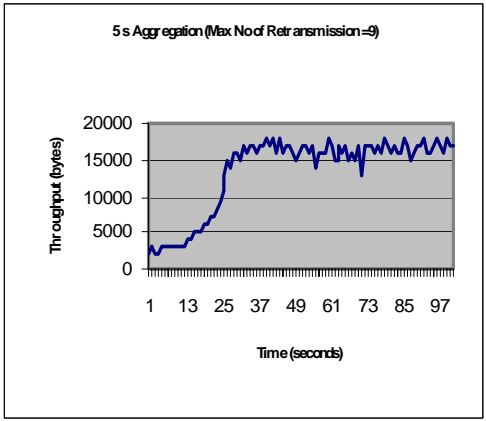
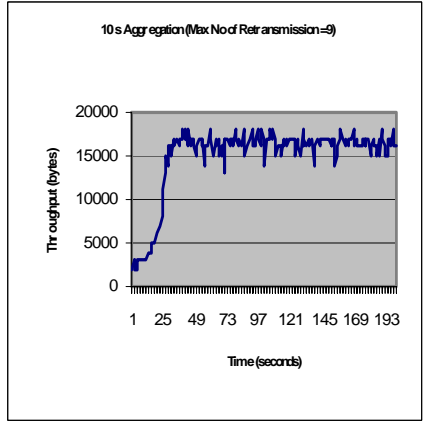
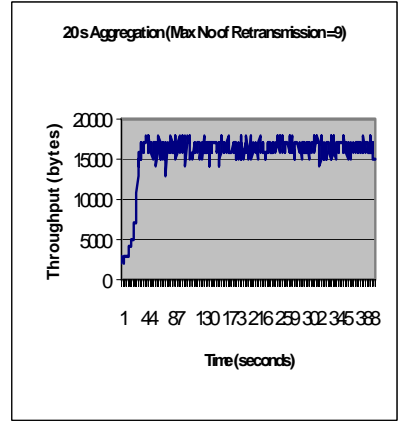
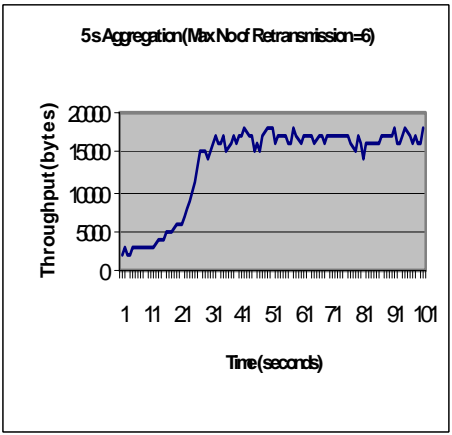
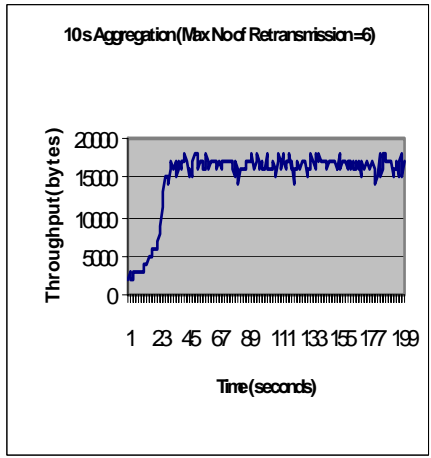
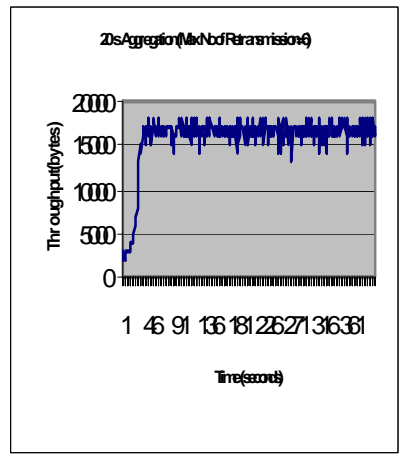
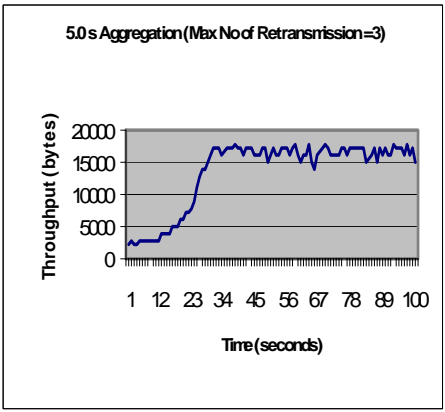
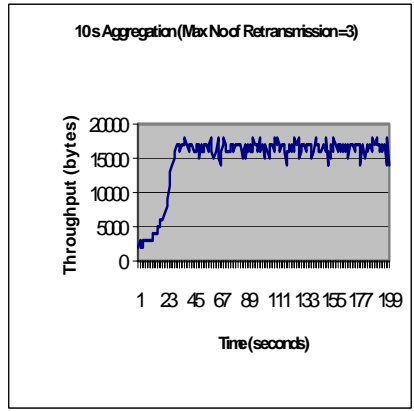
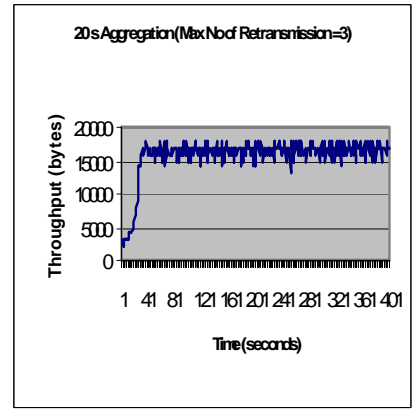

Figure 2 : plots which span three orders of magnitude in time scale and three different ranges of maximum number of retransmission.

## IV. Conclusion

In this paper, the effect of CSMA/CD Exponential Backoff retransmission algorithm on Traffic Self similarity is considered. For this purpose, traffic was investigated under three orders of magnitude in time scale and three different ranges of maximum number of retransmission.

## Acknowledgements

We would like to thank Dr. Moawia El faki for providing support .